# Array Configuration-Agnostic Personalized Speech Enhancement using Long-Short-Term Spatial Coherence

Yicheng Hsu, Yonghan Lee, and Mingsian R. Bai, *Senior Member, IEEE*

*Abstract*—Personalized speech enhancement (PSE) has been a field of active research for suppression of speech-like interferers such as competing speakers or TV dialogues. Compared with single-channel approaches, multichannel PSE systems can be more effective in adverse acoustic conditions by leveraging the spatial information in microphone signals. However, the implementation of multichannel PSEs to accommodate a wide range of array topology in household applications can be challenging. To develop an array configuration-agnostic PSE system, we define a spatial feature termed the long-short-term spatial coherence (LSTSC) as the input feature to a convolutional recurrent network (CRN) to monitor the voice activity of the target speaker. As another refinement, an equivalent rectangular bandwidth (ERB)–scaled LSTSC feature can be used to reduce the computational cost. Experiments were conducted to compare the proposed PSE systems, including the complete and the simplified versions with two baselines using unseen room responses and array configurations (geometry and channel count) in the presence of TV noise and competing speakers. The results demonstrated that the proposed multichannel PSE network trained with the LSTSC feature achieved superior enhancement performance without precise knowledge of the array configurations and room responses.

## I. INTRODUCTION

TELECONFERENCING has become essential amid the COVID-19 pandemic, where working and learning from home has become a norm. However, to enhance the target speech signal when teleconferencing in the presence of speech-like interference such as an interfering dialogue from a television or competing speaker is generally challenging. Such enhancement problem can sometimes be addressed using source separation approaches by which the target signal is extracted from the separated source signals.

Many source separation algorithms were developed in the time–frequency (TF) domain [1]–[5] based on the prior knowledge of the number of sources which can be difficult to estimate accurately in real-world scenarios. In addition, exhaustive separation approaches can be computationally expensive and superfluous because often only the target source is of interest.

Instead of exhaustive separation, a personalized speech enhancement (PSE) system seeks to extract only the target speech signal with auxiliary information in the face of competing speakers or interferers and noise. Various auxiliary information can be derived from video imagery [6]–[9], pre-enrolled utterances [10]–[14], electroencephalogram signals [15], and speech activities [16] pertaining to the target speaker. Although the monaural PSE approaches are effective for close-talking target sources, noticeable signal distortions can be audible for far-field applications such as hands-free teleconferencing and smart speakers. Enhancement performance can degrade due to interference, reverberation, and noise. In these adverse acoustical conditions, multichannel approaches can be more advantageous than their single-microphone counterparts because of the additional spatial information provided by the microphone array. For example, the direction-aware SpeakerBeam [17] employs a self-attention network in a beamformer. The neural spatial filter [18] exploits directional information to extract the target speech. The time-domain SpeakerBeam (TD-SpeakerBeam) [19] utilizes interchannel phase differences (IPDs) as input features to a network for improved separation performance. Some studies suggested direct feature generation based on multichannel microphone signals by using a trainable spatial encoder rather than the ad hoc spatial features [20, 21]. As a limitation of these methods, the microphone array used in the testing phase must be identical to the one used in the training phase. To overcome the limitation, an array geometry-agnostic PSE model [22] was suggested by introducing an ad hoc stream pooling layer to perform multichannel PSE for arbitrary microphone configurations. In addition, the multichannel PSE is conducted in the short-time Fourier transform (STFT) domain, which is computationally expensive.

In this paper, we propose an array configuration-agnostic PSE (ARCA-PSE) system based on a convolution recurrent network (CRN) with spatial features and pre-enrolled utterances as inputs. A novel spatial feature termed long-short-

This work was supported by the National Science and Technology Council (NSTC), Taiwan, under the project number 110-2221-E-007-027-MY3. (*Corresponding author: Mingsian R. Bai*).

Yicheng Hsu is with the Department of Power Mechanical Engineering, National Tsing Hua University, Hsinchu, Taiwan (e-mail: shane.ychsu@gmail.com).

Yonghan Lee was with the Department of Power Mechanical Engineering, National Tsing Hua University, Hsinchu, Taiwan. She is now with RelaJet Tech Co., Ltd., Taipei, Taiwan (e-mail: melinda@relajet.com).

Mingsian R. Bai is with the Department of Power Mechanical Engineering and Electrical Engineering, National Tsing Hua University, Hsinchu, Taiwan (e-mail: msbai@pme.nthu.edu.tw).

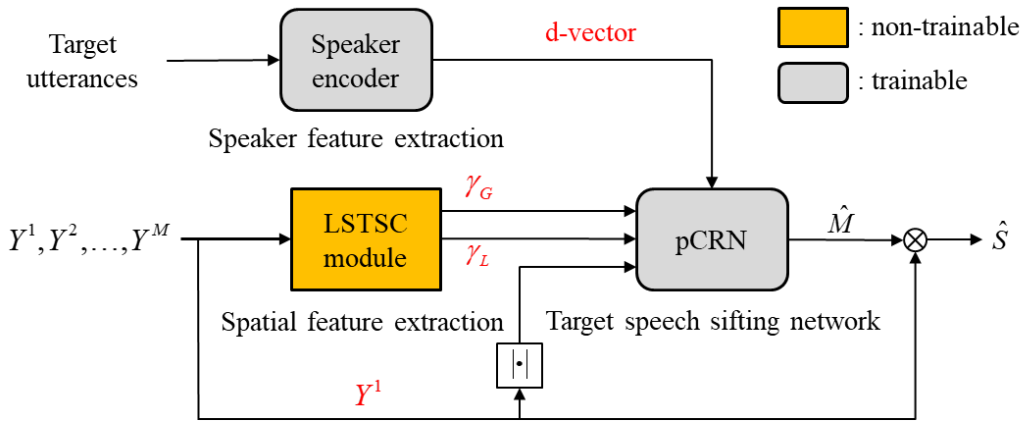

**Fig. 1.** Block diagram of the proposed ARCA-PSE system.

term spatial coherence (LSTSC) associated with each time-frequency (TF) bin is computed in relation to the target speaker activity. The proposed learning-based PSE system proves effective, even for "mismatched" room impulse responses (RIRs) and array configurations (array geometry, inter-element spacing, and number of microphones) which are not the same as those used in network training. The proposed LSTSC feature enables to differentiate the static interferer from the target speaker on the basis of the coherence between the whitened long-term relative transfer function (RTF) and the short-term instantaneous RTF. To prevent the long-term spatial features from mis-converging to the target speaker's RTF, recursive averaging is halted when the target speaker is present. Instead of processing signals in the STFT domain as in the conventional approaches, the proposed LSTSC feature is represented in the equivalent rectangular bandwidth (ERB) filterbank [23].

Unlike the network models reported in [22, 24] that made use of simulated room impulse responses (RIRs) in both training and testing phases, our models employ simulated RIRs in the training phase, but a combination of simulated and recorded RIRs in the testing phase. To assess the robustness of the proposed approach, two experiments are conducted. First, performance improvement attained using the proposed LSTSC feature versus the conventional IPD feature are examined when unseen array geometries are involved in the testing phase. Second, the impact of different number of microphones (including the single-microphone case) on the enhancement performance is examined. DNS-MOS p.835 [25], scale-invariant signal-to-distortion ratio (SI-SDR) [26], and short-time objective intelligibility (STOI) [27] are adopted as the performance measures.

The main contributions of this paper are threefold. First, an array configuration-agnostic PSE system which is robust to unseen RIRs and array configurations is developed. Second, LSTSC is proposed to effectively capture the spatial feature of the target speaker. Third, the computational complexity of LSTSC is reduced considerably using the ERB filterbank.

The remainder of this paper is organized as follows. Section II presents the proposed ARCA-PSE system. Section III discusses the experimental results of the proposed system in comparison with the baselines. Conclusions are given in Section IV.

## II. PROPOSED ARCA-PSE SYSTEM

### A. Problem formulation and signal model

In this section, we propose an ARCA-PSE system that extracts the target speech signal from multichannel mixture signals, with the aid of a target speaker's embedding. Consider a scenario in which the target and non-target speaker's utterances are captured using $M$ distant microphones in the presence of speech-like noise from a television. Speech utterances of the target and non-target speakers at unknown but fixed locations are generated spontaneously. The TV noise represents an interference that is persistent in time and fixed in space. Here, we assume that no prior knowledge of the array configuration is available. The array signal model can be expressed in the STFT domain, with $l$ and $f$ denoting the time frame index and the frequency bin index, respectively. The signal captured by the $m$th microphone can be written as

$$Y^m(l,f) = \sum_{j=1}^{2} A_j^m(f) S_j(l,f) + A_n^m(f) N(l,f) + V^m(l,f), \quad (1)$$

where $A_j^m(f)$ denotes the acoustic transfer function (ATF) between the $m$th microphone and the $j$th speaker, $A_n^m(f)$ denotes the ATF between the $m$th microphone and the interferer, $S_1(l,f)$, $S_2(l,f)$, and $N(l,f)$ denote the STFTs of the target signal, the non-target speaker, and the TV interferer, and $V^m(l,f)$ denotes the additive sensor noise.

### B. Problem formulation and signal model

To tackle the enhancement problem stated above, we proposed an ARCA-PSE system to extract the target speaker signals from the noisy mixture signals. Figure 1 illustrates the overall system architecture composed of three trainable and non-trainable modules: the spatial feature extraction module (Section C) based on LSTSC, the speaker embedding extraction module (Section D) based on the pre-enrolled target speaker utterances, and the target speech sifting network based on the auxiliary information provided by the two preceding modules and a personalized CRN (pCRN) (Section E). All modules will be detailed in the sequel.

*C. The LSTSC feature*

In this study, we propose a novel spatial feature termed the long-short term spatial coherence (LSTSC) to distinguish the speaker source that can be arbitrarily situated from the fixed interferer. It will be demonstrated that LSTSC is a very effective spatial feature applicable to all array geometries and sensor counts and is therefore array configuration-agnostic. For each TF bin, the short-term RTF between the mth microphone and reference microphone 1 is calculated by averaging $(2R+1)$ frames:

$$\tilde{R}^m(l,f) \equiv \frac{\hat{\Phi}_{y^m y^1}}{\hat{\Phi}_{y^1 y^1}} \equiv \frac{\sum_{n=l-R}^{l+R} Y^m(n,f) Y^{1*}(n,f)}{\sum_{n=l-R}^{l+R} Y^1(n,f) Y^{1*}(n,f)} \quad (2)$$

where * denotes the complex conjugate operation, $\hat{\Phi}_{y^m y^1}$ denotes the short-term cross-spectral density estimate between channels $m$ and 1, and $\hat{\Phi}_{y^1 y^1}$ denotes the short-term autospectral density of the reference microphone. Next, a "whitened" RTF vector $\mathbf{r}(l,f) \in \mathbb{R}^{M-1}$, with the first entry (which is always one) deleted, can be used as a short-term spatial feature:

$$\mathbf{r}(l,f) = \left[ \frac{\tilde{R}^2(l,f)}{|\tilde{R}^2(l,f)|}, \ldots, \frac{\tilde{R}^M(l,f)}{|\tilde{R}^M(l,f)|} \right]^T \quad (3)$$

where $|\cdot|$ denotes the complex modulus.

For spatially stationary interferers, the following long-term RTF is computed via recursive averaging:

$$\bar{r}^m(l,f) = \lambda \bar{r}^m(l-1,f) + (1-\lambda) r^m(l,f), \quad m=2,\ldots,M, \quad (4)$$

where $0 < \lambda < 1$ is the forgetting factor that regulates the averaging process. The larger the factor is, the smoother the time-frame average becomes. The long-term RTF vector $\bar{\mathbf{r}}(l,f)$ is also whitened to serve as a long-term spatial feature vector:

$$\bar{\mathbf{r}}(l,f) = \left[ \frac{\bar{r}^2(l,f)}{|\bar{r}^2(l,f)|}, \ldots, \frac{\bar{r}^M(l,f)}{|\bar{r}^M(l,f)|} \right]^T. \quad (5)$$

To exploit the temporal–spatial coherence conveyed by the whitened RTFs, we can calculate the LSTSC, $\gamma(l,f)$, between the short-term whitened feature vector $\mathbf{r}(l,f)$ and the long-term whitened feature vector $\bar{\mathbf{r}}(l,f)$ as follows:

$$\gamma(l,f) = \frac{\operatorname{Re}\{\mathbf{r}^H(l,f) \bar{\mathbf{r}}(l,f)\}}{\sqrt{\mathbf{r}^H(l,f)\mathbf{r}(l,f)} \sqrt{\bar{\mathbf{r}}^H(l,f)\bar{\mathbf{r}}(l,f)}}, \quad (6)$$

$$= \frac{1}{M-1} \operatorname{Re}\{\mathbf{r}^H(l,f) \bar{\mathbf{r}}(l,f)\}$$

where Re{·} denotes the real-part operator and the ($M$-1) in the denominator results from the fact that the feature vectors have been whitened. Here, a complex inner product of the short-term RTF and the long-term RTF is computed, which can also be regarded as a sign-sensitive cosine similarity based on the Euclidean angle [28]. The LSTSC, $-1 \leq \gamma(l,f) \leq 1$, is an indicator of the coherence between the short-term and the long-term spatial features. Furthermore, LSTSC is a scalar feature independent of the number of microphones. To be specific, different numbers of microphones would only affect feature vector dimensions in Eqs. (3) and (5) rather than the input dimension in Eq. (6). In this work, we define the LSTSC with a large $\lambda$ as the global LSTSC ($\gamma_G$), whereas the LSTSC with a small $\lambda$ are defined as the local LSTSC ($\gamma_L$). The larger $\gamma_G$ is close to 1, the more likely the current TF bin pertains to the TV interference. Otherwise, it is likely for either the target speaker and the non-target speaker be present or silence period to occur. In addition, the larger $\gamma_L$ is close to 1, the more likely the current TF bin contains the directional sources. It can be seen in Fig. 2 that $\gamma_G$ tends to sift out the TF bins related to the active speaker or to both the speaker and the interference which are rendered inactive, whereas $\gamma_L$ are used to identify the TF bins corresponding to the directional sources.

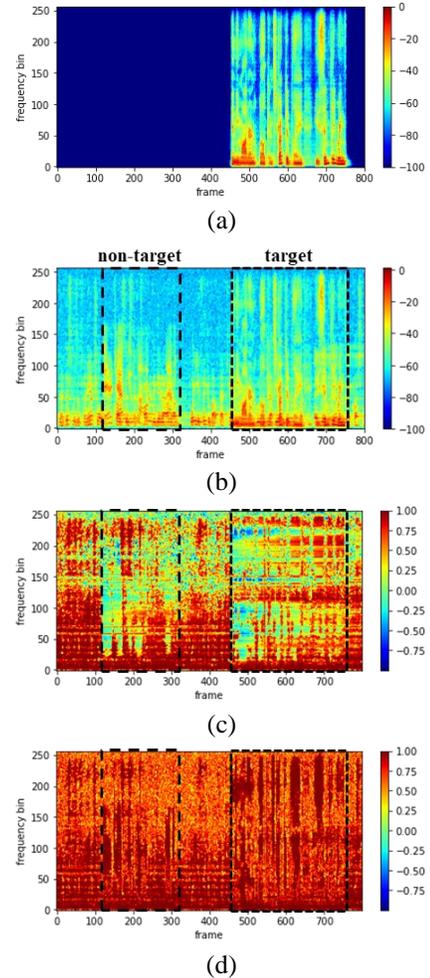

**Fig. 2.** Example spectrograms of the LSTSC features calculated with different forgetting factors. (a) Target speech signal, (b) noisy mixture signal, (c) the global LSTSC ($\lambda$ = 0.99), and (d) the local LSTSC ($\lambda$ = 0.01).

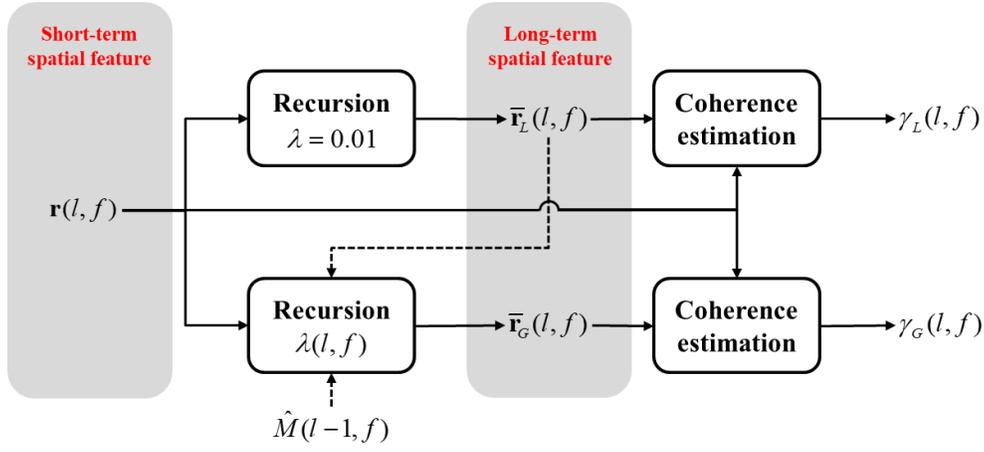

**Fig. 3.** The flowchart to compute the global and the local LSTSCs.

As mentioned previously, $\gamma_G$ tends to sift out the target and non-target-dominant TF bins. However, recursive averaging of the long-term spatial feature vector in Eq. (4) should be halted in the presence of the target signals to prevent the long-term spatial feature vector from mis-converging to the target speaker's RTF. To remedy this problem, we introduce the global LSTSC ($\gamma_G$) with a time-varying forgetting factor, and the detailed procedure to compute the modified global LSTSC ($\gamma_G$) is depicted in Fig. 3.

The long-term spatial feature vector $\bar{\mathbf{r}}_G(l,f)$ of the global LSTSC, $\gamma_G$, is calculated using the following update equation modified from Eq. (4):

$$\bar{r}_G^m(l,f) = \lambda(l,f)\bar{r}_G^m(l-1,f) + [1-\lambda(l,f)]r^m(l,f), \quad (7)$$

$$\lambda(l,f) = \begin{cases} 1 & \text{if } \frac{1}{F}\sum_{f=1}^{F}|\hat{M}(l-1,f)|^2 > \beta \\ 1 - \frac{\gamma_L(l,f)}{20}, & \text{otherwise} \end{cases}, \quad (8)$$

where $\beta$ is a threshold that is set to 0.01 in this paper, $\hat{M}(l-1,f)$ is the soft mask estimated by the pCRN (Fig. 1). With the time-varying forgetting factor $\lambda(l,f)$, the recursive process will be halted in two situations: one is when the target speaker is active, and the other is when no directional source is detected ($\gamma_L$ is small).

$$\gamma'(l,f) = 1 - \frac{2}{\pi}\cos^{-1}[\gamma(l,f)] = \frac{2}{\pi}\sin^{-1}[\gamma(l,f)], \quad (9)$$

Note that $-1 \leq \gamma'(l,f) \leq 1$.

The aforementioned LSTSC feature is calculated in the STFT domain. Considerable computational saving can be gained in light of the equivalent rectangular bandwidth (ERB) [23]. The spectrum of a signal $Y(l,f)$ can be represented in a small number of $B$ bands:

$$Y_{ERB}(l,b) = \sum_{f \in \{f_{b1},...,f_{bF_b}\}} w_b(f)|Y(l,f)|^2, \, b \in \{0,1,...,B\}, \quad (10)$$

$$\gamma_{ERB}(l,b) = \frac{1}{\pi_b}\sum_{f \in \{f_{b1},...,f_{bF_b}\}} w_b(f)\gamma(l,f), \, b \in \{0,1,...,B\}, \quad (11)$$

where $w_b(f)$ and $F_b$ are the weight and the number of the frequency bins for band $b$, respectively. The weight normalization factor

$$\pi_b = \sum_{f}^{F_b} w_b(f), \quad (12)$$

With this ERB approach, the dimension of the LSTSC feature can be reduced to $B$ bands ($B = 48$ in this paper).

Several example spectrograms of the global LSTSC feature are shown in Fig. 4. The global LSTSC with $\lambda = 0.99$ is shown in Fig. 4(c), where the presence of target speech is incorrectly identified because the long-term feature $\bar{\mathbf{r}}(l,f)$ mis-converges to target speaker's RTF (the dashed box). On the contrary, the LSTSC feature with a time-varying forgetting factor does not suffer from this problem and effectively detect the presence of target speech, as shown in Fig. 4(d). From Fig. 4(e), the LSTSC feature with inverse sine can further increase the spatial resolution in the low-frequency range. To reduce the computational cost, the ERB-scaled LSTSC feature is computed to detect the target speaker activity with only 48 bands, as shown in Fig. 4(f).

*D. Speaker encoder*

The proposed learning-based PSE system also requires a speaker embedding produced by a speaker encoder based on the pre-enrolled utterances of the target speaker. The speaker encoder is jointly trained using either the target speech model or a pre-trained model to generate speaker embeddings in the forms of the i-vector [29], x-vector [30], or d-vector [31]. In this study, we used the d-vector which has been widely used in speaker diarization, speech synthesis, speech translation, personal voice activity detection, source separation, voice preservation tests, etc. The speaker encoder is comprised of a three-layer long short-term memory (LSTM) network and is trained using a generalized end-to-end loss function suggested in a previous study [31]. The VoxCeleb2 data set was used in the training phase [32]. The speaker embeddings are output in sliding windows. The resulting aggregated embeddings, or the d-vector, of the target speaker serve as the input features to the pCRN module.

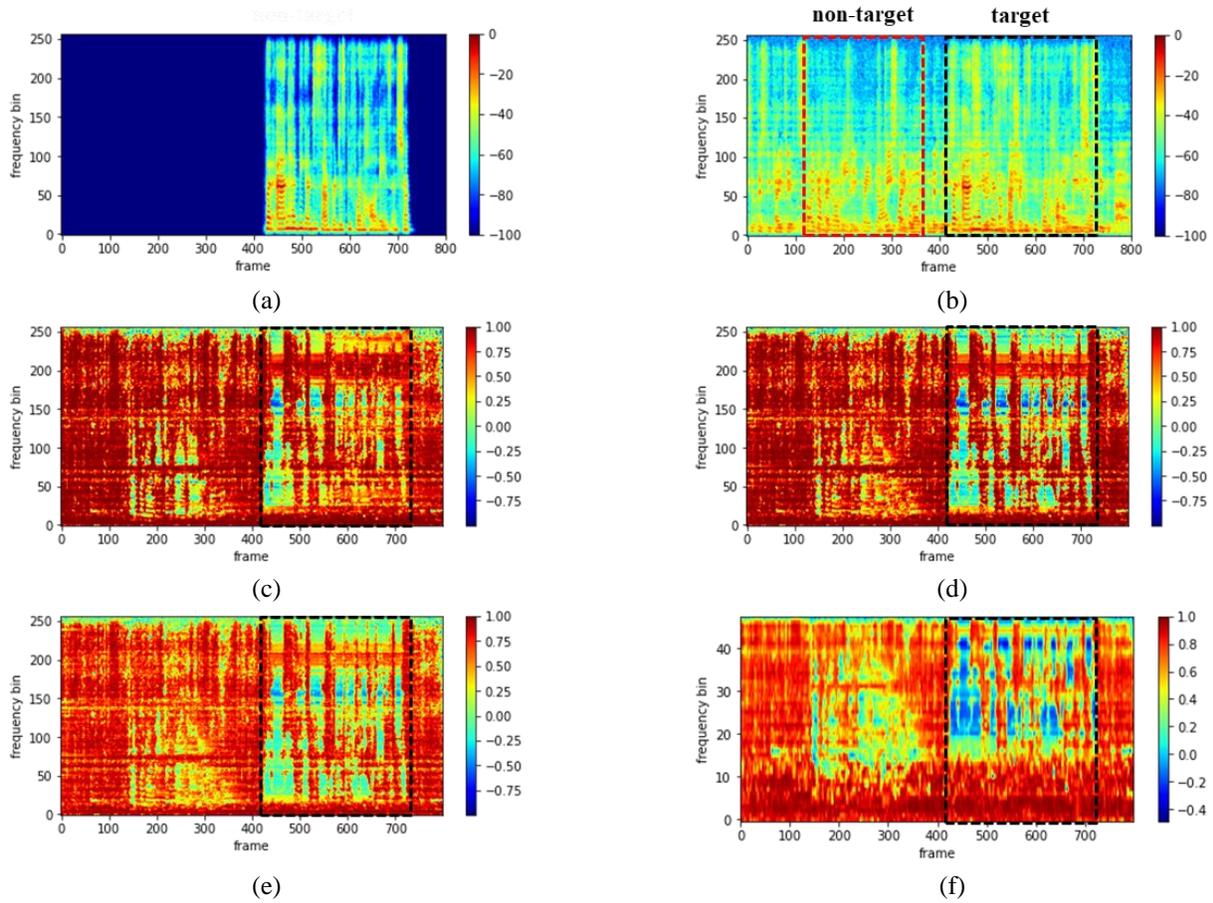

**Fig. 4.** Example spectrogram of the LSTSC feature: (a) target speech signal, (b) noisy mixture signal, (c) global LSTSC ($\lambda = 0.99$), (d) global LSTSC with adaptive $\lambda$, (e) global LSTSC with adaptive $\lambda$ + arcsine transformation, and (f) ERB-scaled global LSTSC with adaptive $\lambda$ + arcsine transformation, where the forgetting factor is modified into a time-varying parameter.

*E. Target speech sifting network*

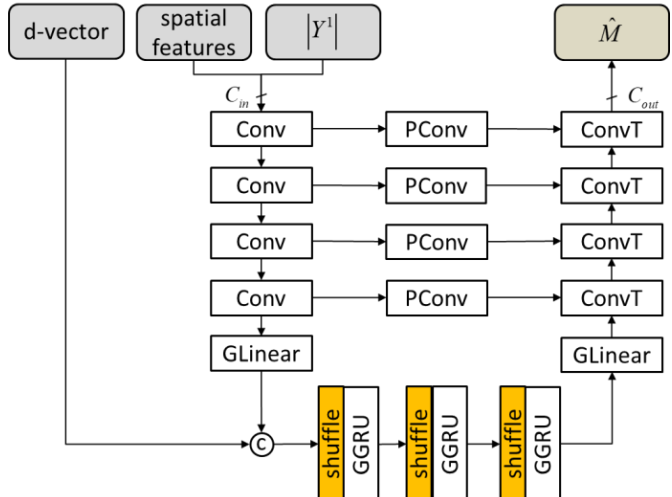

**Fig. 5.** The target speech sifting network.

The target speech sifting network, pCRN, is based on the CRN and U-Net [33]. In Fig. 5, the pCRN admits three inputs, including the magnitude spectrogram of the reference microphone signal, the LSTSC feature, and the d-vector generated by the speaker encoder. In this network, spatially stationary and temporally persistent interferences are suppressed by the encoder layers and the grouped linear layer. Next, the d-vector is concatenated with the output of the grouped linear (GLinear) layer for every time frame. The resulting vector is shuffled and then fed to the following grouped gated recurrent unit (GGRU) layers to sift latent features associated with the target speaker. The noisy spectrum of the reference microphone signal is multiplied element-wise with the soft mask generated by the pCRN to yield the enhanced spectrum. As illustrated in Fig. 3, the pCRN has four symmetric encoder and decoder layers with a 16-32-64-128 filter. The input channels $C_{in}$ depends on the number of the input feature channel, and the output channel $C_{out}$ is 1 in this study. The convolutional blocks consist of a separable convolution, followed by batch normalization and ReLU activation, whereas the output layer uses sigmoid activation. The convolution kernel and step size are set to (3,2) and (2,1). Instead of concatenating the encoder output with the corresponding decoder input, 1×1 pathway convolutions (PConv) are employed as add-skip connections, which leads to considerable parameter reduction with little performance degradation [34]. Computational complexity of the linear and GRU layer is reduced further by using the grouping technique [34, 35, 36]. The input layer is subdivided into $P$ groups, resulting in $P$ smaller linear/GRU layers. Here, $P = 4$ in the GLinear and GGRU layers. The input to each grouped GRU (GGRU) layer is shuffled to recover the inter-group correlations. Three GGRU layers with 256 nodes in each are adopted.

## III. EXPERIMENTAL STUDY

Experiments were conducted to validate the proposed ARCA-PSE system. The network is trained using simulated RIRs, but tested using both simulated and measured RIRs.

*A. Data preparation*

Clean utterances which contain utterances from 921 and 40 speakers were selected from the train-clean-360 and dev-clean subsets of the LibriSpeech corpus [37] for training and testing. Noisy speech signals including 74-hour human conversation clips in Youtube® from the VoxConverse data set [38] were used for training and testing. The audio contains noise of various types such as conversation, background noise, music, laughter, and applause. The experiment was conducted at a sample rate of 16 kHz.

*B. Training and validation dataset*

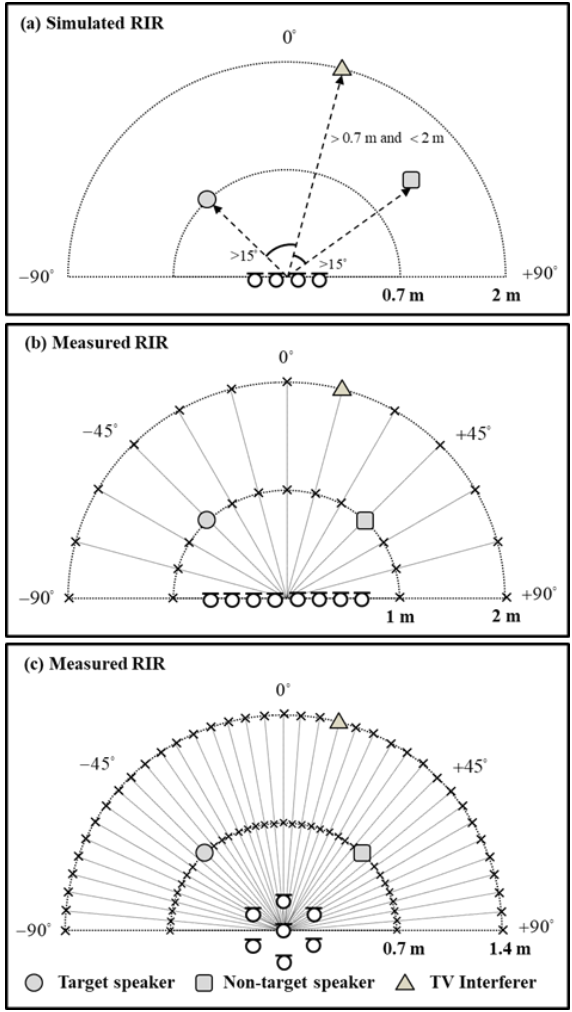

**Fig. 6.** Illustration of experimental settings for different RIRs.

In total, 50,000 and 5,000 samples are used in training and validation. Noisy audio signals edited in eight-second clips were prepared by mixing the target speech signal and speech-like TV noise with signal to interference ratio (SIR) = 0, 5, 10, and 15 dB. The target speech signal and the non-target speech signal are selected to be equal in power. In addition, sensor noise was added with signal-to-noise ratio (SNR) = 20, 25, and 30 dB. The target and non-target speaker signals were not overlapped in time. As illustrated in Fig. 6(a), the target speaker, speech-like TV interference, and another interfering non-target speaker were randomly placed in the frontal plane in the ring sector bounded by radius = 0.7 m and 2 m. The target speaker is set to be closer to the array center than the non-target speaker and the TV interferer. In addition, any two sources are at least 15° apart from each other. A four-element uniform linear array (ULA) with inter-element spacing of 8 cm is used in training and validation. Reverberant microphone signals were simulated by convolving the clean signals with RIRs by using the image-source method [39], with the reverberation time ($T_{60}$) = 0.1, 0.3, 0.5, and 0.7 s.

*C. Test set*

We created three 2,000-sample test datasets to assess the robustness of the proposed ARCA-PSE system when applied to unseen RIRs, unseen array geometries, and unknown number of microphones. The experimental settings of the first dataset are the same as those used in training and validation, as depicted in Fig. 6(a), with the only difference being that $T_{60}$ is set to 0.16, 0.36, and 0.61 s. The second dataset was created by using the measured RIRs in the Multi-Channel Impulse Responses Database [40], which recorded using an eight-element ULA with an inter-element spacing of 8 cm for $T_{60}$ = 0.16 s, 0.36 s, and 0.61 s at Bar-Ilan University. The RIRs were measured in 15° intervals from −90° to 90° at a distance of 1 m and 2 m from the array center, as depicted in Fig. 6(b). On the other hand, we created another 2,000 seven-channel mixture signals based on the recently proposed Tampere University Rotated Circular Array Dataset [41] recorded using a seven-element microphone array, with the one microphone at the center and the other six fitted on a circle of 8 cm diameter. The RIRs were measured from −90° to 90° in 5° intervals, in a distance of 0.7 m and 1.4 m from the array center, as illustrated in Fig. 6(c).

*D. Baseline methods and implementation detail*

Two baseline models were used to compare with the proposed ARCA-PSE system. All the models were based on pCRN architecture (Fig. 2). The first baseline was a pCRN model with no spatial input feature. The second baseline was a multichannel pCRN (MC-pCRN) model with IPDs as auxiliary spatial feature. For the proposed ARCA-PSE system, the MC-pCRN model with four LSTSC features were implemented. The settings of the LSTSC features for the proposed methods are detailed in Table I. Table II summaries the parameter size and the multiply-accumulate operations (MACs) for the baseline models as well as the proposed models. The result suggests that the LSTSC implementation in the ERB domain requires the minimal memory and computation costs.

To train the baseline and the proposed models, the signal frame was selected to be 25 ms in length with stride 10 ms. A 512-point FFT was used. The optimizer was Adam with a learning rate of 0.001 and a gradient norm clipping of 3. The learning rate was halved if the loss of the validation set failed to improve for three consecutive epochs.

TABLE I
THE SETTINGS FOR THE LSTSC FEATURE

|  | $\lambda_L$ | $\lambda_G$ | Inverse sine | ERB-scaled |
|---|---|---|---|---|
| LSTSC-1 | 0.01 | 0.99 |  |  |
| LSTSC-2 | 0.01 | Time-varing |  |  |
| LSTSC-3 | 0.01 | Time-varing | v |  |
| LSTSC-4 | 0.01 | Time-varing | v | v |

TABLE II
NUMBER OF PARAMETERS AND MACs FOR THE PROPOSED APPROACHES AND FOR THE BASELINES

| Model | Params [M] | MACs [M] |
|---|---|---|
| pCRN | 1.01 | 4.68 |
| MC-pCRN w/ IPD | 1.01 | 4.75 |
| MC-pCRN w/ LSTSC-1 | 1.01 | 4.72 |
| MC-pCRN w/ LSTSC-2 | 1.01 | 4.72 |
| MC-pCRN w/ LSTSC-3 | 1.01 | 4.72 |
| MC-pCRN w/ LSTSC-4 | **0.79** | **2.02** |

*E. Evaluation metrics*

In this study, several subjective and objective metrics were used to evaluate PSE performance. DNS-MOS P. 835 [24] provides subjective evaluation scores, including background noise quality (BAK), signal quality (SIG), and overall audio quality (OVRL) for the evaluation of deep learning models. We also used STOI [26] to measure speech intelligibility. DNS-MOS and STOI were calculated only for the period when the target speech was present. In addition, another objective metric SI-SDR [25] was used to evaluate overall enhancement quality regardless of target speech absence or presence.

*F. Results and discussions*

Microphones with various array configurations are adopted in the comparison of the proposed model and the baselines for different array geometries and RIRs, as shown in Fig. 7(a). Table III summaries the enhancement performance of all approaches. The results show that the multichannel PSE systems based on LSTSC and IPDs performed comparably when the array configurations and RIRs identical to the settings in the training phase. However, if the RIRs are "unseen" in the testing phase, significant performance degradation can be observed for the system based on IPDs, even though the same array configuration was used as in the training set. In addition, if the array geometry is different from that used in training, performance drops drastically. In contrast, the proposed PSE system based on LSTSC still performs satisfactorily despite the unseen RIRs and array geometries. The enhancement performance of the proposed system further improves when the adaptive parameter ($\lambda$) and the inverse sine function are used in calculating LSTSC. In addition, tremendous computational saving is achieved by leveraging the ERB-scaled at the cost of minor enhancement performance degradation. One example obtained using the proposed PSE system based on LSTSC is shown in Fig. 8. The simulation is conducted using measured RIRs with T60 = 320 ms and array geometry, G3. It can be observed from the spectrograms that the PSE system based on LSTSC was able to extract reliably the target speech (the dashed box), even if the array configuration and room response are unseen in the testing phase.

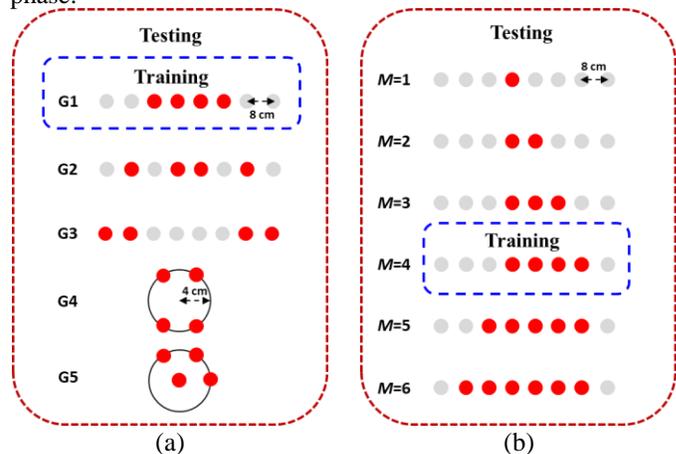

Fig. 7. Microphone array settings for simulations to investigate the effects of (a) array configurations and (b) number of microphones.

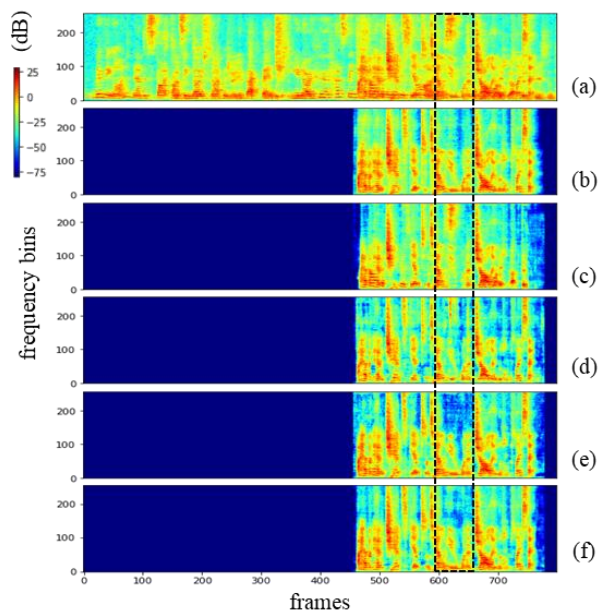

**Fig. 8.** Example spectrograms of a test sample with 5 dB SNR (TV noise) and 0 dB SIR (competing speaker). (a) Noisy signal, (b) target speech signal, and the outputs of (c) pCRN w/o spatial information, (d) MC-pCRN w/ IPD feature, (e) MC-pCRN w/ LSTSC-3, and (f) MC-pCRN w/ LSTSC-4.

TABLE III
COMPARISONS OF ENHANCEMENT PERFORMANCE IN TERMS OF DNS-MOS, SISDR, AND STOI FOR DIFFERENT ARRAY GEOMETRIES

| Array Geometry | Method | BAK | SIG | OVLR | SI-SDR | STOI |
|---|---|---|---|---|---|---|
| G1 w/ simulated RIRs | Noisy | 1.71 | **3.58** | 1.85 | -1.76 | 78.44 |
| | pCRN | 3.43 | 3.16 | 2.42 | 9.06 | 84.15 |
| | MC-pCRN w/ IPD | 3.48 | 3.23 | 2.51 | 11.57 | 90.28 |
| | MC-pCRN w/ LSTSC-1 | **3.82** | 3.18 | 2.51 | 11.24 | 89.95 |
| | MC-pCRN w/ LSTSC-2 | 2.92 | 3.48 | 2.57 | 11.74 | 91.17 |
| | MC-pCRN w/ LSTSC-3 | 3.04 | 3.53 | **2.67** | **12.01** | **91.44** |
| | MC-pCRN w/ LSTSC-4 | 3.41 | 3.37 | 2.63 | 11.15 | 90.23 |
| G1 w/ measured RIRs | Noisy | 1.74 | **3.60** | 1.91 | -1.29 | 79.91 |
| | pCRN | 3.75 | 3.08 | 2.51 | 8.80 | 83.96 |
| | MC-pCRN w/ IPD | 3.61 | 2.89 | 2.34 | 5.79 | 85.39 |
| | MC-pCRN w/ LSTSC-1 | **4.00** | 3.07 | 2.50 | 11.17 | 90.13 |
| | MC-pCRN w/ LSTSC-2 | 3.42 | 3.39 | 2.78 | 11.80 | 92.18 |
| | MC-pCRN w/ LSTSC-3 | 3.47 | 3.48 | **2.87** | **12.28** | **92.42** |
| | MC-pCRN w/ LSTSC-4 | 3.78 | 3.34 | 2.82 | 11.82 | 91.39 |
| G2 w/ measured RIRs | Noisy | 1.74 | **3.60** | 1.91 | -1.29 | 79.91 |
| | pCRN | 3.75 | 3.08 | 2.51 | 8.80 | 83.96 |
| | MC-pCRN w/ IPD | 3.90 | 2.57 | 2.13 | 1.33 | 71.62 |
| | MC-pCRN w/ LSTSC-1 | **4.00** | 3.07 | 2.49 | 11.27 | 90.26 |
| | MC-pCRN w/ LSTSC-2 | 3.43 | 3.39 | 2.79 | 11.86 | 92.38 |
| | MC-pCRN w/ LSTSC-3 | 3.45 | 3.46 | **2.85** | **12.38** | **92.60** |
| | MC-pCRN w/ LSTSC-4 | 3.76 | 3.28 | 2.81 | 11.90 | 91.51 |
| G3 w/ measured RIRs | Noisy | 1.74 | **3.60** | 1.91 | -1.29 | 79.91 |
| | pCRN | 3.75 | 3.08 | 2.51 | 8.80 | 83.96 |
| | MC-pCRN w/ IPD | 3.88 | 2.53 | 2.08 | 0.54 | 70.52 |
| | MC-pCRN w/ LSTSC-1 | **4.00** | 3.06 | 2.49 | 11.31 | 90.49 |
| | MC-pCRN w/ LSTSC-2 | 3.42 | 3.40 | 2.79 | 11.92 | 92.56 |
| | MC-pCRN w/ LSTSC-3 | 3.46 | 3.48 | **2.87** | **12.47** | **92.73** |
| | MC-pCRN w/ LSTSC-4 | 3.73 | 3.33 | 2.80 | 11.96 | 91.70 |
| G4 w/ measured RIRs | Noisy | 1.93 | **3.57** | 2.05 | -0.92 | 81.16 |
| | pCRN | 3.52 | 2.98 | 2.35 | 7.94 | 80.13 |
| | MC-pCRN w/ IPD | 3.83 | 2.71 | 2.20 | -0.72 | 71.21 |
| | MC-pCRN w/ LSTSC-1 | **3.84** | 3.06 | 2.43 | 11.70 | 87.37 |
| | MC-pCRN w/ LSTSC-2 | 3.26 | 3.31 | 2.69 | 13.57 | 91.80 |
| | MC-pCRN w/ LSTSC-3 | 3.35 | 3.39 | **2.78** | **14.33** | **92.01** |
| | MC-pCRN w/ LSTSC-4 | 3.61 | 3.29 | 2.70 | 13.90 | 91.12 |
| G5 w/ measured RIRs | Noisy | 1.93 | **3.57** | 2.05 | -0.92 | 81.16 |
| | pCRN | 3.52 | 2.98 | 2.35 | 7.94 | 80.13 |
| | MC-pCRN w/ IPD | 3.84 | 2.67 | 2.16 | -0.78 | 70.55 |
| | MC-pCRN w/ LSTSC-1 | **3.86** | 3.07 | 2.44 | 12.00 | 87.62 |
| | MC-pCRN w/ LSTSC-2 | 3.27 | 3.32 | 2.70 | 13.92 | 91.95 |
| | MC-pCRN w/ LSTSC-3 | 3.36 | 3.37 | **2.78** | **14.99** | **92.11** |
| | MC-pCRN w/ LSTSC-4 | 3.64 | 3.28 | 2.71 | 14.38 | 91.10 |

Next, we examine the effect of the sensor count on speech enhancement. The IPD-based network trained with a one-array configuration cannot be applied to the test data set gathered using another configuration because the IPD feature is already fixed for a particular array geometry and number of microphones. However, for the proposed LSTSC feature, a change in the number of microphones only results in a change in feature vector dimensions in Eqs. (3) and (5) and not the input dimension in Eq. (6). Different number of microphones are examined, as shown in Fig. 7(b). Moreover, an additional single-microphone system was included in the experiment to benchmark the proposed approach. As indicated in results summarized in Table IV, the enhancement performance in both bin-wise and band-wise pCRN were significantly improved in terms of DNS-MOS, SI-SDR, and STOI with an increase in the number of microphones, although we only trained using a four-element array. Without the requirement of changes in model architecture for individual arrays, a single model trained with the proposed LSTSC feature can be shared between multiple arrays with different shapes and the number of microphones. The single-microphone system, representing the approach with no spatial information, yielded the worst performance. Figure 9 shows the spectrograms estimated by the multichannel PSE system based on LSTSC (time-varying $\lambda$ and inverse sine) for different numbers of microphones. The result shows that the more microphones are used, the better estimate can be achieved.

TABLE IV
COMPARISONS OF ENHANCEMENT PERFORMANCE IN TERMS OF DNS-MOS, SISDR, AND STOI FOR DIFFERENT NUMBERS OF MICROPHONES

| Method | Mic. number | BAK | SIG | OVLR | SI-SDR | STOI |
|---|---|---|---|---|---|---|
| pCRN | 1 mic | **3.70** | 3.13 | 2.56 | 9.17 | 84.67 |
| MC-pCRN w/ LSTSC-3 | 2 mics | 3.22 | 3.41 | 2.71 | 9.57 | 88.72 |
| | 3 mics | 3.34 | 3.55 | 2.85 | 12.07 | 91.93 |
| | 4 mics | 3.42 | 3.56 | 2.90 | 12.47 | 92.44 |
| | 5 mics | 3.46 | 3.57 | 2.92 | 12.63 | 92.63 |
| | 6 mics | 3.50 | **3.57** | **2.95** | **12.77** | **92.80** |
| ERB-scaled pCRN | 1 mic | 3.72 | 3.07 | 2.52 | 8.96 | 83.63 |
| MC-pCRN w/ LSTSC-4 | 2 mics | 3.70 | 3.28 | 2.73 | 10.44 | 88.85 |
| | 3 mics | 3.74 | 3.39 | 2.84 | 11.72 | 90.86 |
| | 4 mics | 3.76 | 3.41 | 2.87 | 12.00 | 91.41 |
| | 5 mics | 3.77 | 3.43 | 2.88 | 12.12 | 91.65 |
| | 6 mics | **3.79** | 3.43 | 2.89 | 12.23 | 91.86 |

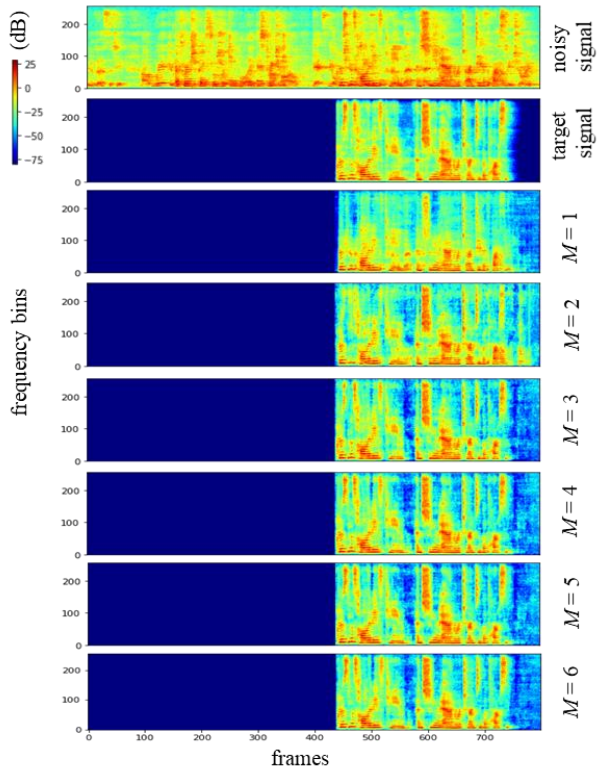

Fig. 9. Signal spectrograms (TV noise at 5 dB SNR and competing speaker interference at 0 dB) of noisy signal, target signal, and the enhanced spectrograms obtained by 1-6 microphones, respectively.

## IV. CONCLUSIONS

In this work, we presented an ARCA-PSE system based on LSTSC in conjunction with the target speaker embedding. The results demonstrated that the LSTSC feature enables robust enhancement for unseen RIRs and unseen array configurations, which is highly desirable for real-world applications. In addition, LSTSC with an adaptive forgetting factor and inverse sine function can further improve the enhancement performance. The ERB-scaled LSTSC effectively reduced the computational complexity with only minor performance loss, making it amenable to embedded systems.